\documentclass{aastex63}
\begin{document}
\title{Relative Ages of Nine Inner Milky Way Globular Clusters from Proper Motion Cleaned Color-Magnitude Diagrams\footnote{Based on
    observations made with the NASA/ESA Hubble Space Telescope, obtained at the
    Space Telescope Science Institute, which is operated by the Association of
    Universities for Research in Astronomy, Inc., under NASA contract NAS
    5-26555.  These observations are associated with program GO-15065.}} 

\correspondingauthor{Roger E. Cohen}
\email{rcohen@stsci.edu}

\author{Roger E. Cohen}
\affiliation{Space Telescope Science Institute, 3700 San Martin
  Drive, Baltimore, MD 21218, USA}

\author{Andrea Bellini}
\affiliation{Space Telescope Science Institute, 3700 San Martin
  Drive, Baltimore, MD 21218, USA}

\author{Luca Casagrande}
\affiliation{Research School of Astronomy and Astrophysics, Australian National University, Australia}

\author{Thomas M. Brown}
\affiliation{Space Telescope Science Institute, 3700 San Martin
  Drive, Baltimore, MD 21218, USA}

\author{Matteo Correnti}
\affiliation{Space Telescope Science Institute, 3700 San Martin Drive, Baltimore, MD 21218, USA}

\author{Jason S. Kalirai}
\affiliation{Johns Hopkins University Applied Physics Laboratory, 11101 Johns Hopkins Road, Laurel, MD 21723, USA}

\begin{abstract}

Our picture of the age-metallicity relation for Milky Way globular clusters (MWGCs) is still highly incomplete, and the majority of MWGCs lack self-consistent age measurements.  Here, we exploit deep, homogenous multi-epoch \textit{Hubble Space Telescope} (\textit{HST}) imaging of nine MWGCs located towards the inner Milky Way to measure their relative ages, in most cases for the first time.  Our relative 
age measurements are designed to be directly comparable to the large set of MWGC ages presented by \citealp*{v13} (\citeauthor{v13}), 
using identical filters, evolutionary models, and bolometric corrections, extended to the higher extinction values relevant to our target clusters.  Adopting the \citeauthor{v13} MWGC age scale, our relative age measurements imply that 
our target clusters are consistently very old, with a mean age of 12.9$\pm$0.4 Gyr, with the exception of the young metal-rich MWGC NGC 6342.  We perform two tests to validate the precision of our methodology, and discuss the implications of our target cluster loci in the MWGC age-metallicity plane.  In addition, we use our fully self-consistent bolometric corrections to assess the systematic impact of variations in the total-to-selective extinction ratio $R_{V}$ on relative age measurements.  

\end{abstract}


\section{Introduction} \label{sec:intro}

The ages and metallicities 
of 
MWGCs pose crucial constraints on Milky Way formation and evolution.  By using MWGCs as tracers, substantial progress towards understanding the assembly history of the Milky Way has been made possible by the combination of homogenous iron abundances (\citealp*{c09}, hereafter \citeauthor{c09}, their Table A.1) and self-consistent age measurements for large ensembles of MWGCs from high quality photometry (\citealt{vandenberg2000}, \citealt{salarisweiss}, \citealt{deangeli05}, \citealt{mf09}, \citealt{dotter10}, \citealp*{v13}, hereafter \citeauthor{v13}, \citealt{omalley17}). 
The resulting age-metallicity relation (AMR) revealed a bifurcation \citep{mf09,fb10,leaman13} implying that a substantial fraction of MWGCs were accreted early in the history of the Milky Way, while many others likely formed in situ, and this scenario was reinforced by then-available kinematics \citep[e.g.][]{leaman13}.

A leap forward in linking MWGC properties to the global picture of Milky Way formation has been made by combining age and metallicity information with cluster kinematic properties.
This recent progress is due in large part to the quality of space-based MWGC absolute proper motions \citep[e.g.][]{sohn}, in particular from the all-sky \textit{Gaia} mission \citep{gaia,gaiaedr3}.  These absolute proper motions, when combined with 
distances and a model of the Milky Way potential, yield orbital parameters \citep{baumgardt19,massari19,baumgardt21}.
Using the additional information available from orbital criteria, individual MWGCs can be linked to a specific progenitor accreted onto the Milky Way, or categorized as being formed in situ 
\citep{myeong18,massari19}.  Furthermore, such information can be combined with state-of-the-art simulations to ascertain which parameters are most useful to constrain the accretion history of the Milky Way \citep{kruijssen_sims}.  By matching MWGCs to their progenitors, information is gained not only about each cluster individually, but about the formation history of the Milky Way via the nature (mass and time of accretion) of the progenitors themselves \citep{kruijssen_gc1,kruijssen_gc2}.

While the power of a large ensemble of MWGCs to unlock key events in Milky Way formation and evolution is clear, the full discriminatory power of MWGCs in the inner Milky Way (where a substantial fraction of its MWGC population actually resides) has not been realized.  Most of these clusters have been excluded from our current picture of the MWGC age-metallicity relation simply because they lack age estimates and, in many cases, reliable metallicity estimates.  As an additional complication, distances from literature compilations, used to calculate orbital properties, can be unreliable as well, placing clusters on the wrong side of the Galactic bulge in some cases \citep[e.g.][]{javiervar,cadelano6256}.  In addition, we know that not all MWGCs presently located in the inner Milky Way are confined to the Galactic bulge, instead showing a kinematic diversity including interlopers from the halo and thick disk 
\citep{ortolani_intruders,cadelano6256,perezvillegas}.  Therefore, placing inner Milky Way globulars on the MWGC age-metallicity relation is critical not only to determine progenitors or discriminate interlopers on a case-by-case basis, but also to constrain the global fraction of present-day MWGCs that were formed in situ \citep[e.g.][]{keller,trujillogomez}.

Here, we exploit multi-epoch, homogenous space-based imaging of an ensemble of nine inner Milky Way globular clusters to measure their ages.  We have already demonstrated the potential of this dataset from both a photometric and astrometric perspective in \citet{dyn}, where we analyzed radial density and proper motion dispersion profiles of our target clusters, finding that they are preferentially highly dynamically evolved.  Here, we use the proper-motion-cleaned color-magnitude diagrams (CMDs) to gain additional insight on the evolutionary history of these clusters to place them on the MWGC age-metallicity relation in the context of their more well-studied counterparts from \citet{ataggc}.  Importantly, because systematics can play a significant role in MWGC age estimates (see e.g.~the Appendix of \citealt{massari19}), we conduct our analysis self-consistently with the oft-cited MWGC age investigation of \citeauthor{v13} from both observational and theoretical standpoints, using an identical observing setup as well as identical models, including self-consistent bolometric corrections, to obtain our MWGC age estimates.

The remainder of this paper is organized as follows: In the next section, we summarize the observations and data reduction used to produce differential-reddening-corrected, proper-motion-cleaned photometric catalogs.  In Sect.~\ref{relagesect} we describe the technique used to measure cluster ages, and our results are discussed in Sect.~\ref{discusssect} along with some outstanding issues.  In the final section, we present our conclusions.

\section{Data \label{datasect}} 

Our nine target clusters were selected as having extant archival HST imaging of sufficient depth and spatial resolution to reach several magnitudes faintward of the main sequence turnoff in the cluster cores.  To measure relative proper motions, we obtained a second epoch of deep imaging with ACS/WFC onboard HST (GO-15065, PI:Cohen), using the F606W and F814W filters to produce homogenous photometry with an identical observing setup as that used for the ACS Galactic Globular Cluster Treasury Survey \citep{ataggc}, on which several recent MWGC age compilations are based (\citealt{mf09}, \citealt{dotter10}, \citeauthor{v13}).  The details of both epochs of observations for all target clusters, separated by time baselines of $\gtrsim$9 yr, are given in Table 1 of \citet{dyn}\footnote{The MAST data used in this work are available at \dataset[https://doi.org/10.17909/t9-na33-8504]{10.17909/t9-na33-8504}}.

The tools and techniques we used to produce astrophotometric catalogs including PSF photometry and relative proper motions for our target clusters are identical to those used in other recent studies \citep{bellini14,bellini_wcen_phot,bellini_pm,libralato_pm}, and further details, including photometric quality cuts and extensive artificial star tests, are given in \citet{dyn}.  The resulting catalogs yield proper-motion-cleaned, Vegamag-calibrated PSF photometry extending several magnitudes faintward of the main sequence turnoff (MSTO) of all target clusters, and as far brightward as the red giant branch (RGB), and in most cases the horizontal branch (HB), limited only by the archival first epoch observations\footnote{While the brightest cluster giants near the tip of the RGB are saturated in one or both epochs, the saturation limit occurs significantly ($>$1 mag in all cases) brightward of the magnitude range needed for our relative age analysis described in Sect.~\ref{techniquesect}.}.  We have calculated differential reddening corrections using proper-motion-selected members as described elsewhere \citep[e.g.][]{milone12,bellini_wcen_dr}, and the photometry we analyze below consists of proper-motion-selected cluster members, with differential reddening corrections applied.

\section{Relative Age Measurements \label{relagesect}}

\subsection{Models \label{modsect}}

To enable a direct comparison with the 
age measurements of \citeauthor{v13}, we use identical stellar evolutionary tracks and the publicly available software from \citet{vrisos12} and \citet{vrisos14} to compute theoretical isochrones from the evolutionary tracks.  We assume [$\alpha$/Fe]=+0.4 since high-resolution spectroscopy reveals inner Milky Way globular clusters to be enhanced in $\alpha$-elements (\citealt{cesar6440,sandrom28,barbuy6558,johnson6569}; also see table 7 of \citealt{j16} and references therein), and assume a helium abundance of Y=0.25 for [Fe/H]$\leq$1.0, slightly increasing towards higher [Fe/H] as $\Delta$Y/$\Delta$Z$\sim$1.4 following \citeauthor{v13}.  
To convert these isochrones to the observational plane, we have calculated bolometric corrections in the filters relevant to our observations (ACS/WFC F606W and F814W) identically as described in \citet{cv14}, but extended to higher values of $E(B-V)$$\leq$ 3 beyond the maximum value of $E(B-V)$=0.72 available through their public software.  
These bolometric corrections are computed using \texttt{MARCS} synthetic spectra \citep{marcs} and assume the extinction law of \citet{ccm} and \citet{odonnell} (the impact of variations in the properties of interstellar extinction is discussed further in Sect.~\ref{rvsect}).

\subsection{Fiducial Sequences \label{fidsect}}

For each target cluster, we construct a fiducial sequence from the differential-reddening-corrected, proper-motion-cleaned photometry in the (F606W-F814W),F814W plane (since we fit color as a function of magnitude, the subgiant branch is less perpendicular to the fitting axis than if we had used the F606W filter on the vertical CMD axis).   

\begin{figure}
\gridline{\fig{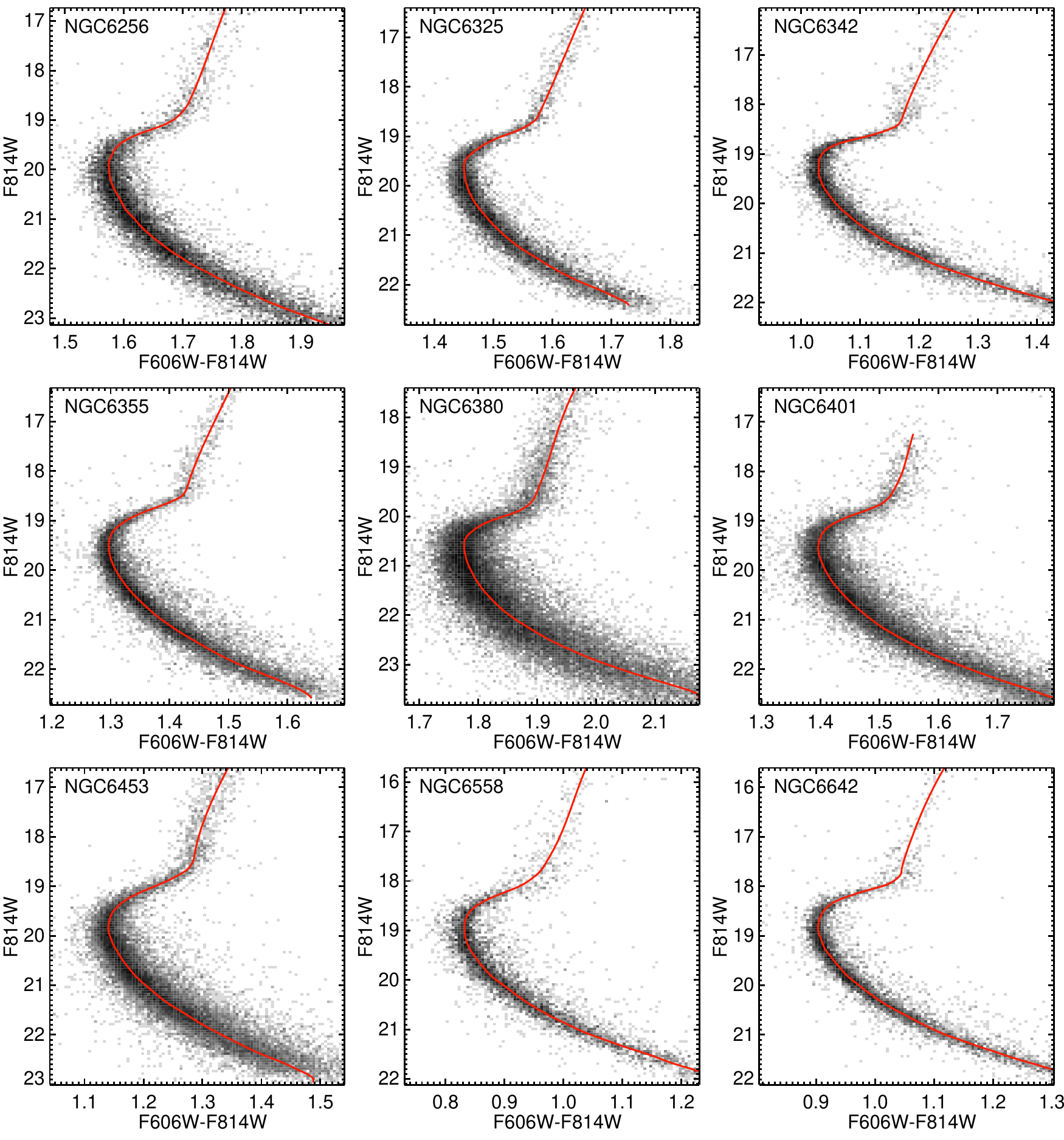}{0.85\textwidth}{}}      	 
\caption{Proper motion cleaned, differential reddening corrected photometric catalogs of our target clusters in the region near the MSTO, shown as Hess diagrams (color-magnitude density plots, Log-scaled in this case). 
Fiducial sequences are shown in red.
\label{targfidfig}}
\end{figure}

The procedure we use to generate a fiducial sequence is quite similar to those used in the recent literature \citep[i.e.][]{mf09,rachel_lmc}: The median color is calculated in 0.1 mag bins (rotated to be parallel to the fiducial sequence from the previous iteration if not the first iteration) using a 2.5$\sigma$ clip to reject outliers (i.e.~blue stragglers and high-mass-ratio binaries), and the RGB is fit with a 3rd order polynomial\footnote{The base of the RGB is defined as the maximum in the second derivative of color with respect to magnitude.}.  This procedure is repeated 10 times, shifting the magnitude bins by one-tenth of the binsize (0.01 mag) each time.  The 10 resulting fiducial sequences are averaged, and this process is repeated several times (typically three, with a maximum of five) until convergence is indicated by a sum of absolute deviations from the previous iteration of $<$0.001 mag.  To quantify the uncertainties on this fiducial sequence, we repeat the entire procedure for 10,000 monte carlo iterations, where several sources of uncertainty are injected in each monte carlo iteration: First, for each observed star, we apply color and magnitude offsets randomly drawn from the nearest (in color, magnitude and distance from the cluster center) artificial stars\footnote{Specifically, for each observed star, artificial stars are drawn from those within 0.4 mag in F814W, 0.2 mag in F606W-F814W (using their raw observed magnitudes before differential reddening correction), and 20\% in distance from the cluster center, giving thousands of artificial stars per observed star.}.  By drawing directly from an ensemble of artificial stars for each observed star, we require no a priori assumptions about whether photometric error distributions are Gaussian, have null offsets (bias), or whether color and magnitude errors are correlated.  In addition, differential reddening uncertainties are taken into account by offsetting each star along the reddening vector by Gaussian deviates of the uncertainty on its differential reddening correction, 
and lastly, an offset in each filter is applied drawn from the (potentially correlated) distribution of photometric zeropoint uncertainties.
By sampling from error distributions of color, magnitude, zeropoint, and differential reddening correction over each of the monte carlo iterations, we obtain a fiducial sequence in each iteration, which is interpolated to a fixed magnitude grid.  To account for any potential photometric bias, the final fiducial sequence color at each magnitude point is the median color over the monte carlo iterations, and the uncertainty in the fiducial sequence color is the 16th-84th percentile interval over the monte carlo iterations.  Fiducial sequences for our target clusters are shown in Fig.~\ref{targfidfig}.

\subsection{Relative Age Measurement Technique \label{techniquesect}}

\begin{figure}
\gridline{\fig{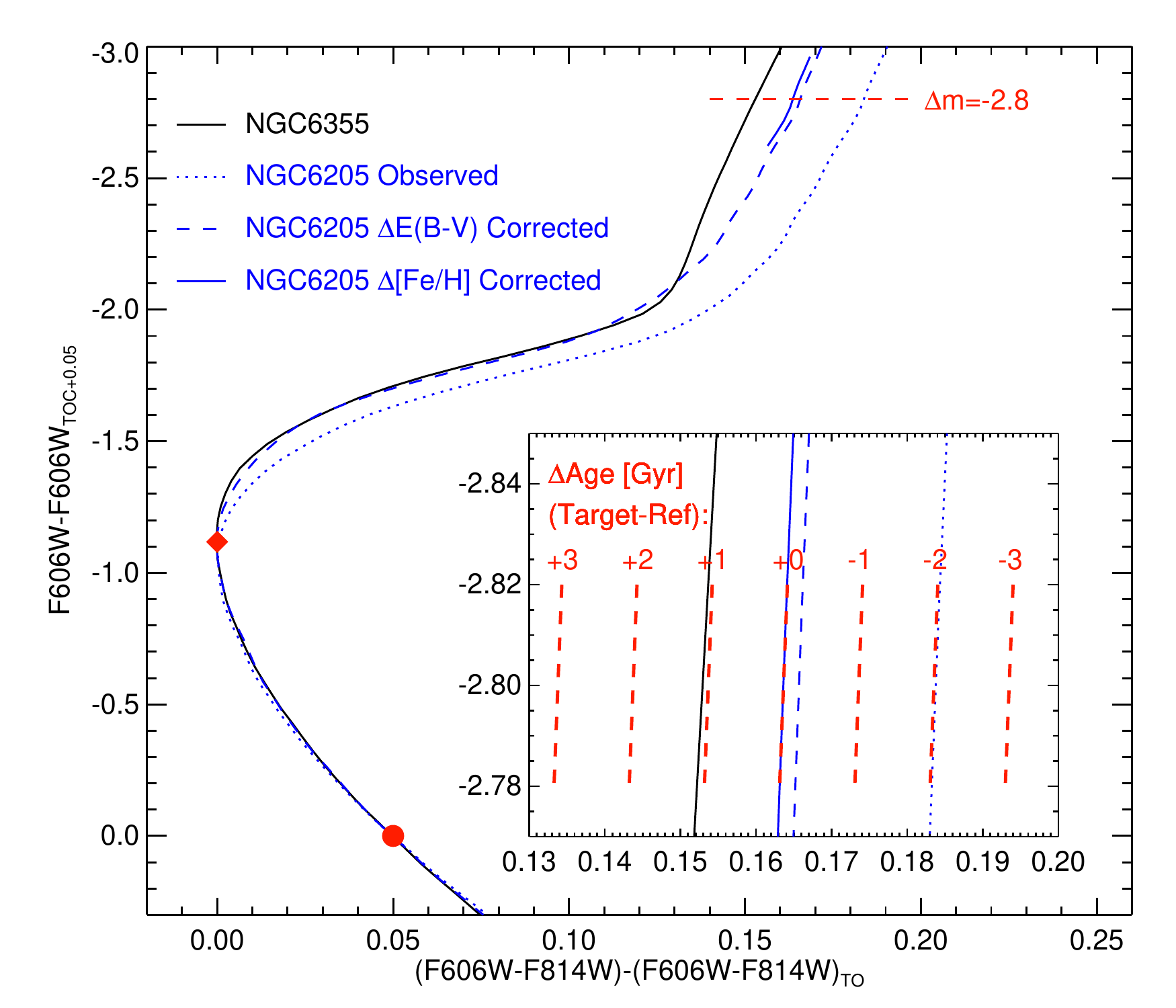}{0.7\textwidth}{}}      	 
\caption{An example of the CMD registration procedure used to measure relative ages.  In this case the reference cluster, NGC 6205 (= M13), is being registered to the target cluster NGC 6355.  The target cluster fiducial sequence is shown as a black line.  The observed reference cluster fiducial sequence (dotted blue line) has its shape corrected for the difference in $E(B-V)$ between the target and reference clusters (dashed blue line).  The two fiducial sequences are then matched in color at their MSTO (red diamond) and in magnitude at the point on the main sequence 0.05 mag redward of the MSTO (red circle).  Lastly, an isochrone-based correction to the RGB color (solid blue line) is applied to account for any difference in [Fe/H] between the target and reference clusters. The relative age is then directly measured from the color difference between the RGB and MSTO (F606W-F814W)-(F606W-F814W)$_{\rm TO}$ on the lower RGB at a magnitude of F606W-F606W$_{\rm TOC+0.05}$=$-$2.8, shown using a dashed red horizontal line.  In the inset, we show a zoomed-in version of the registered CMD near this RGB magnitude, where the dashed red lines indicate the shift in the RGB color as a function of age relative to the reference cluster predicted by the isochrones, labeled in red. \label{regexamplefig}}
\end{figure}

We measure the relative ages of our target clusters using the classical MSTO-to-RGB color difference as described by \citet{vandenberg90} and implemented by \citeauthor{v13} \citep[also see][]{sd90,chaboyer96,buonanno98}.  
This method is designed to assess the age difference between a target cluster (with unknown age) and a reference cluster (generally with a known age) through a CMD registration procedure, illustrated for an example case in Fig.~\ref{regexamplefig}.  To measure relative ages, the fiducial sequences of both the target and reference cluster are shifted along the horizontal axis so that their MSTO colors coincide, setting the zeropoint of the horizontal axis in the registered CMD, shown as a red diamond in Fig.~\ref{regexamplefig}.  Next, the fiducial sequences are shifted vertically so that their magnitudes coincide at the magnitude on the main sequence (i.e.~faintward of the MSTO) where the fiducial sequence color is 0.05 mag redder than the MSTO color\footnote{Measuring the magnitude of a fiducial sequence 0.05 redward of the MSTO \citep{buonanno98} is much easier than attempting to measure the MSTO magnitude, since the fiducial sequence is, by definition, vertical at the MSTO.} and this magnitude is set as the new zeropoint of the vertical axis in the registered CMD, shown as a red circle in Fig.~\ref{regexamplefig}.  The reason for registering the fiducial sequences in this way, as discussed extensively by e.g.~\citet{vandenberg90} and \citeauthor{v13}, is that the age difference between the target and reference clusters can now be directly read from the difference in their registered RGB colors at a chosen registered magnitude, for example $-$2.8 mag above the new (registered) magnitude zeropoint, shown as a dashed horizontal red line in Fig.~\ref{regexamplefig}.  In the case where the target and reference clusters do not have identical values of [Fe/H], an isochrone-based correction is applied to the color of the lower RGB.  To illustrate the relative age-dating procedure, the inset of Fig.~\ref{regexamplefig} shows the difference between target and reference cluster age that would be obtained as a function of the color of the lower RGB, illustrating that bluer (redder) registered RGB colors correspond to older (younger) target cluster ages relative to their reference cluster.

Normally, ages and their uncertainties from the above registration procedure are straightforward to calculate because the registration procedure only involves vertical and horizontal shifts of the cluster fiducial sequences in the CMD.   
In our case, application of this technique comes with an additional step, which is needed because our target clusters are typically at substantially higher values of $E(B-V)$ than the well-studied reference clusters to which we compare them.  In particular, because the \textit{shape} of an isochrone changes as a function of total extinction,
two clusters at substantially different values of $E(B-V)$ cannot be compared by simply shifting the fiducial sequence of one to match the other.  Therefore, for each reference cluster, we use the bolometric corrections we have calculated (see Sect.~\ref{modsect}) to correct the reference cluster fiducial sequence for the difference in $E(B-V)$ between the target cluster and the reference cluster.  This correction can change the shape of the reference cluster fiducial sequence quite significantly, as shown by the difference between the dotted and dashed blue lines in Fig.~\ref{regexamplefig}.  Only once the change in isochrone shape due to the difference in $E(B-V)$ has been accounted for do we then apply an isochrone-based correction to the color of the lower RGB to account for any difference in [Fe/H] between target and reference clusters, shown as a solid blue line segment along the RGB in Fig.~\ref{regexamplefig}.  Importantly, this relative age dating procedure, including the corrections for both the $E(B-V)$ difference and [Fe/H] difference between target and reference clusters, does not assume that the isochrones are correct in an absolute sense; rather they are used only to apply \textit{relative} corrections to empirically determined fiducial sequences. 

\subsection{Target Cluster Parameters}

Our relative age measurement technique requires assuming, for both the target and reference clusters, per-cluster values and uncertainties for both the color excess $E(B-V)$ (to correct the isochrone shape for the difference in color excess between target and reference clusters) and the iron abundance [Fe/H] (to correct the color of the RGB for any difference in [Fe/H] between the target and reference clusters).  Regarding assumed [Fe/H] values and uncertainties, we have compiled in Table \ref{targfehtab} recent literature estimates of [Fe/H] for our target clusters from various sources.  These include the compilation by \citeauthor{c09} (see their Table A.1), 
and several other studies that each include multiple target clusters.  These are the CaII study by \citealp*{vasquez} (denoted \citeauthor{vasquez}), low resolution optical spectroscopy by \citealp*{dias} (\citeauthor{dias}), estimates from near-IR photometric indices by \citealp*{valenti10} (\citeauthor{valenti10}), and the values given in the catalog of Harris 1996 (2010 edition, hereafter \citeauthor{h96}).  In the last column, we give additional estimates from other individual sources on a per-cluster basis, where values from high-resolution optical spectroscopy of individual cluster giants, which we deem most trustworthy, are given in boldface and adopted as our metallicity estimates for those clusters (NGC 6342 and NGC 6558).  However, for the majority of our sample, metallicity estimates from high-resolution spectra are notably absent, so we adopt an estimate and uncertainty 
based on the recent literature values in Table \ref{targfehtab}.  For target cluster $E(B-V)$ values, we draw from either \citeauthor{h96} or \citeauthor{valenti10}, averaging them when both are available, and conservatively assume a 15\% uncertainty on $E(B-V)$.

\begin{deluxetable}{lcccccl}
\tablecaption{Literature [Fe/H] Abundances for Target Clusters \label{targfehtab}}
\tablehead{
  \colhead{Cluster} & \colhead{\citeauthor{c09}} & \colhead{\citeauthor{vasquez}\tablenotemark{a}} & \colhead{\citeauthor{dias}} & \colhead{\citeauthor{valenti10}} & \colhead{\citeauthor{h96}} & \colhead{Other\tablenotemark{b}} 
}
\startdata
NGC 6256 & -0.62$\pm$0.09\tablenotemark{c} & -1.61$\pm$0.14 & -1.05$\pm$0.13 & -1.63 & -1.02 & \\
NGC 6325 & -1.37$\pm$0.14 & & & & -1.25 & -1.23$\pm$0.17 (1) \\
NGC 6342 & -0.49$\pm$0.14 & & & -0.71 & -0.55 & -0.60$\pm$0.01 (2); \textbf{-0.53$\pm$0.11 (3)} \\
NGC 6355 & -1.33$\pm$0.14 & -1.59$\pm$0.15 & -1.46$\pm$0.06 & -1.42 & -1.37 & \\
NGC 6380 & -0.40$\pm$0.09\tablenotemark{c} & & & -0.87 & -0.75 & -0.72$\pm$0.17 (1); \textbf{-0.73$\pm$0.06 (4)} \\
NGC 6401 & -1.01$\pm$0.14 & -1.30$\pm$0.15 & -1.12$\pm$0.07 & -1.37 & -0.75 & -1.13$\pm$0.06 (5) \\ 
NGC 6453 & -1.48$\pm$0.14 & -1.67$\pm$0.13 & -1.54$\pm$0.09 & -1.38 & -1.50 & \\
NGC 6558 & -1.37$\pm$0.14 & & -1.01$\pm$0.05 & & -1.32 & -1.07$\pm$0.17 (1); \textbf{-1.17$\pm$0.10 (6)} \\
NGC 6642 & -1.19$\pm$0.14 & & & -1.20 & -1.26 & \\
\enddata
\tablecomments{References: (1)\citet{m14}, using their cubic fit to C09c values (their table 3).  Error is their urms (unweighted rms) of this calibration. (2)\citet{ovr05}  (3)\citet{j16} (4)\citet{jose6380} (5)\citet{tsapras}  (6)\citet{barbuy6558}}
\tablenotetext{a}{Assuming their \citet{saviane} calibration.}
\tablenotetext{b}{Values from high resolution optical spectroscopy of multiple resolved member stars are given in boldface.}
\tablenotetext{c}{For NGC 6256 and NGC 6380, the values given by \citeauthor{c09} result from applying a fixed offset of 0.025 dex to [Fe/H] values from the then-most-recent (2003) edition of the Harris catalog, which are notably different from the more recent \citeauthor{h96} values.}
\end{deluxetable}

\subsection{Reference Clusters}

\begin{figure}
\gridline{\fig{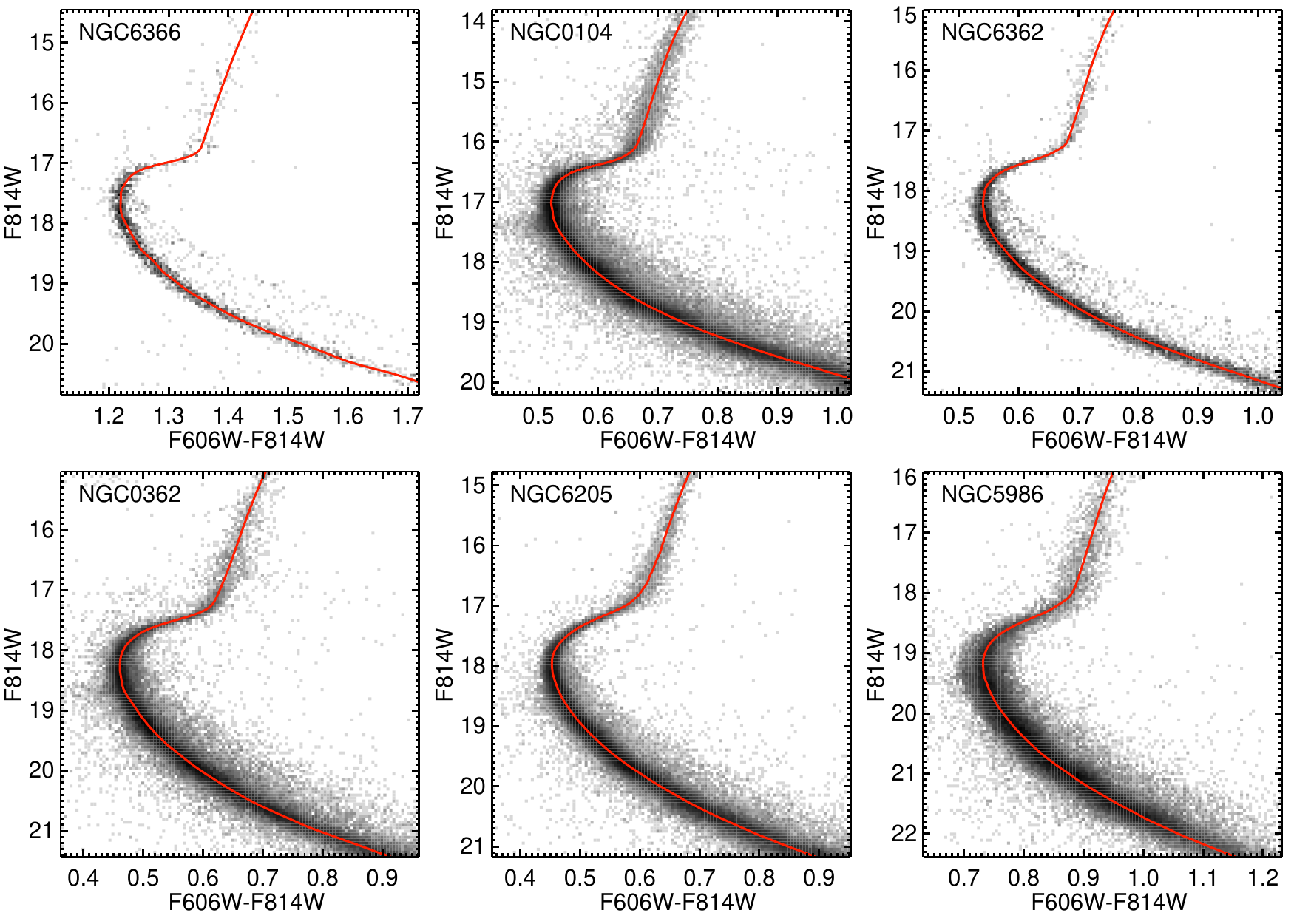}{0.85\textwidth}{}}      	 
\caption{Hess diagrams (color-magnitude density plots, Log-scaled as in Fig.~\ref{targfidfig}) of our reference clusters spanning the metallicity baseline of our target clusters, in order of decreasing [Fe/H].  Of stars in the catalogs of \citet{ataggc}, only those with membership probabilities $p_{\rm mem}>$80\% from \citet{hugs} are included.  
Fiducial sequences are shown in red.
\label{compfidfig}}
\end{figure}

For our reference clusters, we choose MWGCs with ages from \citeauthor{v13} and metallicities similar to (and bracketing the range of) those of our target clusters.  The reference clusters have been intentionally selected to have $\alpha$-enhancement consistent with that assumed for our target clusters \citep{c10,johnson5986,massari6362} and generally low extinction, with the exception of NGC 6366 at the metal-rich end, which high-resolution spectroscopy reveals to be essentially a chemical twin of the target cluster NGC 6342 in terms of [Fe/H], but also O and $\alpha$-elements \citep{j16}.
Reference cluster fiducial sequences and their uncertainties are determined identically as for our target clusters (see Sect.~\ref{fidsect}), using photometric catalogs from the public database of \citet{ataggc} and accompanying artificial star tests, restricted to stars with valid proper motions and membership probabilities $p_{\rm mem}>$80\% from the catalogs of \citet{hugs}.  In addition, due to the substantial impact of differential reddening on the CMD of NGC 6366, differential reddening corrections were applied to this cluster identically as for the target clusters.  
Photometry and fiducial sequences for our reference clusters are shown in Fig.~\ref{compfidfig}.  
The six reference clusters we adopt are listed in Table \ref{refclustab}, along with adopted values and uncertainties of [Fe/H] from \citeauthor{c09} (also used by \citeauthor{v13} to derive ages) and $E(B-V)$ from \citeauthor{v13}, for which we conservatively assume an uncertainty of $\sigma$$E(B-V)$=0.01 mag or 15\%, whichever is larger.

\begin{deluxetable}{cccc}
\tablecaption{Reference Cluster Parameters \label{refclustab}}
\tablehead{
\colhead{Reference Cluster} & \colhead{[Fe/H]} & \colhead{$E(B-V)$} & \colhead{Age} \\ & dex & mag & Gyr  
}
\startdata
NGC6366 & -0.55$\pm$0.10 & 0.730$\pm$0.110 & 11.00$\pm$0.50 \\
NGC0104 & -0.76$\pm$0.02 & 0.032$\pm$0.010 & 11.75$\pm$0.25 \\
NGC6362 & -1.07$\pm$0.05 & 0.076$\pm$0.011 & 12.50$\pm$0.25 \\
NGC0362 & -1.30$\pm$0.04 & 0.032$\pm$0.010 & 10.75$\pm$0.25 \\
NGC6205 & -1.58$\pm$0.04 & 0.017$\pm$0.010 & 12.00$\pm$0.38 \\
NGC5986 & -1.63$\pm$0.08 & 0.280$\pm$0.042 & 12.25$\pm$0.75 \\
\enddata
\tablecomments{Reference cluster [Fe/H] is from \citeauthor{c09}, except NGC 6366, which is from the high resolution spectroscopic study of \citet{j16}.  $E(B-V)$ is from \citeauthor{v13}, and the uncertainties we assume are 15\% or 0.01 mag, whichever is larger.  Relative ages and uncertainties are from \citeauthor{v13}.}
\end{deluxetable}

\subsection{Relative Age Uncertainties}

To calculate the total uncertainties on the relative ages we measure, we take into account several sources of uncertainty, both on the observations as well as the assumed [Fe/H] and $E(B-V)$ for the target and reference clusters, injecting these uncertainties and measuring a relative age for each of the monte carlo iterations in Sect.~\ref{fidsect}:
\begin{itemize}
\item $\sigma_{\rm obs}$ is the uncertainty in registered RGB color due to (potentially correlated) errors in the photometry and photometric zeropoint, described in Sect.~\ref{fidsect}.
\item $\sigma_{E(B-V)}$ is the contribution from the uncertainty in $E(B-V)$, which affects the shape of the reference cluster fiducial sequence. Specifically, an assumed value of $E(B-V)$ is needed to correct the reference cluster fiducial sequence shape for the difference in $E(B-V)$ between the target and reference clusters using our bolometric corrections.
\item $\sigma_{\rm [Fe/H]}$ is the contribution due to the uncertainty of the metallicity correction to the RGB color.  The uncertainties on target and reference cluster [Fe/H] have additional, more subtle contributions to the uncertainty budget because the assumed values of [Fe/H] affect both the gradient of RGB color with relative age (see Sect.~\ref{rvsect}) and the bolometric corrections used to account for a difference in $E(B-V)$ above.
\end{itemize}

The values we assume for $\sigma_{E(B-V)}$ and $\sigma_{\rm [Fe/H]}$ are given in Table \ref{refclustab} for the reference clusters and Table \ref{htab} for our target clusters.
All of the above error sources are injected in each monte carlo iteration to obtain an individual relative age estimate, with the standard deviation of these estimates taken as the uncertainty on the target cluster age \textit{relative to the reference cluster}, listed in Table \ref{htab}.

\begin{deluxetable}{cccccc}
\tablecaption{Target Cluster Relative Ages \label{htab}}
\tablehead{
\colhead{Target Cluster} & \colhead{[Fe/H] (assumed)} & \colhead{$E(B-V)$ (assumed)} & \colhead{Reference Cluster} & \colhead{$\Delta$(Age) (Target-Reference)} & \colhead{Age\tablenotemark{a}} \\
 & dex & mag & & Gyr & Gyr  
}
\startdata
NGC6256 & $-$1.61$\pm$0.20 & 1.15$\pm$0.17 & NGC6205 & 0.8$\pm$0.7 & 12.9$\pm$1.0 \\
NGC6325 & $-$1.37$\pm$0.14 & 0.91$\pm$0.14 & NGC6205 & 0.5$\pm$0.6 & 12.5$\pm$0.9 \\
NGC6342 & $-$0.53$\pm$0.11 & 0.52$\pm$0.08 & NGC6366 & 0.5$\pm$1.0 & 11.5$\pm$1.3 \\
NGC6355 & $-$1.50$\pm$0.20 & 0.79$\pm$0.12 & NGC6205 & 1.1$\pm$0.6 & 13.2$\pm$0.9 \\
NGC6380 & $-$0.73$\pm$0.15 & 1.23$\pm$0.18 & NGC0104 & 1.2$\pm$0.8 & 12.9$\pm$1.1 \\
NGC6401 & $-$1.15$\pm$0.20 & 0.91$\pm$0.14 & NGC6362 & 0.6$\pm$0.9 & 13.2$\pm$1.2 \\
NGC6453 & $-$1.48$\pm$0.14 & 0.67$\pm$0.10 & NGC6205 & 1.2$\pm$0.5 & 13.3$\pm$0.8 \\
NGC6558 & $-$1.17$\pm$0.10 & 0.44$\pm$0.07 & NGC6362 & $-$0.2$\pm$0.8 & 12.3$\pm$1.1 \\
NGC6642 & $-$1.20$\pm$0.15 & 0.50$\pm$0.08 & NGC0362 & 1.9$\pm$0.8 & 12.7$\pm$1.1 \\
\enddata
\tablenotetext{a}{The ages given in the last column are \textit{not} direct absolute age measurements, rather they are calculated based on our relative ages and the comparison cluster absolute ages given by \citeauthor{v13}, see text for details.}
\end{deluxetable}

\subsection{Testing the Method \label{methtestsect}}

We perform two tests as a check on the reliability of our age measurement technique.  First, for each target cluster we use an alternate reference cluster, selecting the next nearest reference cluster in metallicity, which has a metallicity difference opposite in sign to the original reference cluster.  Using this alternate set of reference clusters caused a change in our target cluster ages of $<$0.1 Gyr in the mean, with a standard deviation of $<$0.3 Gyr and an absolute value $<$0.5 Gyr in all cases.  As a second test, we apply our age measurement procedure within our set of reference clusters, treating each one in turn as a target cluster and using the other reference clusters with neighboring [Fe/H] to calculate ages.  We find that we recover the ages listed by \citeauthor{v13} to within their uncertainties across \textit{all} reference-reference cluster pairs, and the difference between the ages we estimate versus those listed by \citeauthor{v13} has a mean of 0.1$\pm$0.2 Gyr.  Given the results of these two tests, the ages we measure for our target clusters should be directly comparable with those given by \citeauthor{v13} at the level of $\sim$0.3 Gyr or better (modulo assumptions on chemical abundances, discussed below in Sect.~\ref{otherchemsect}), well within their individual uncertainties.

\section{Results and Discussion \label{discusssect}}

\subsection{The Age-Metallicity Relation \label{amrsect}}

\begin{figure}
\gridline{\fig{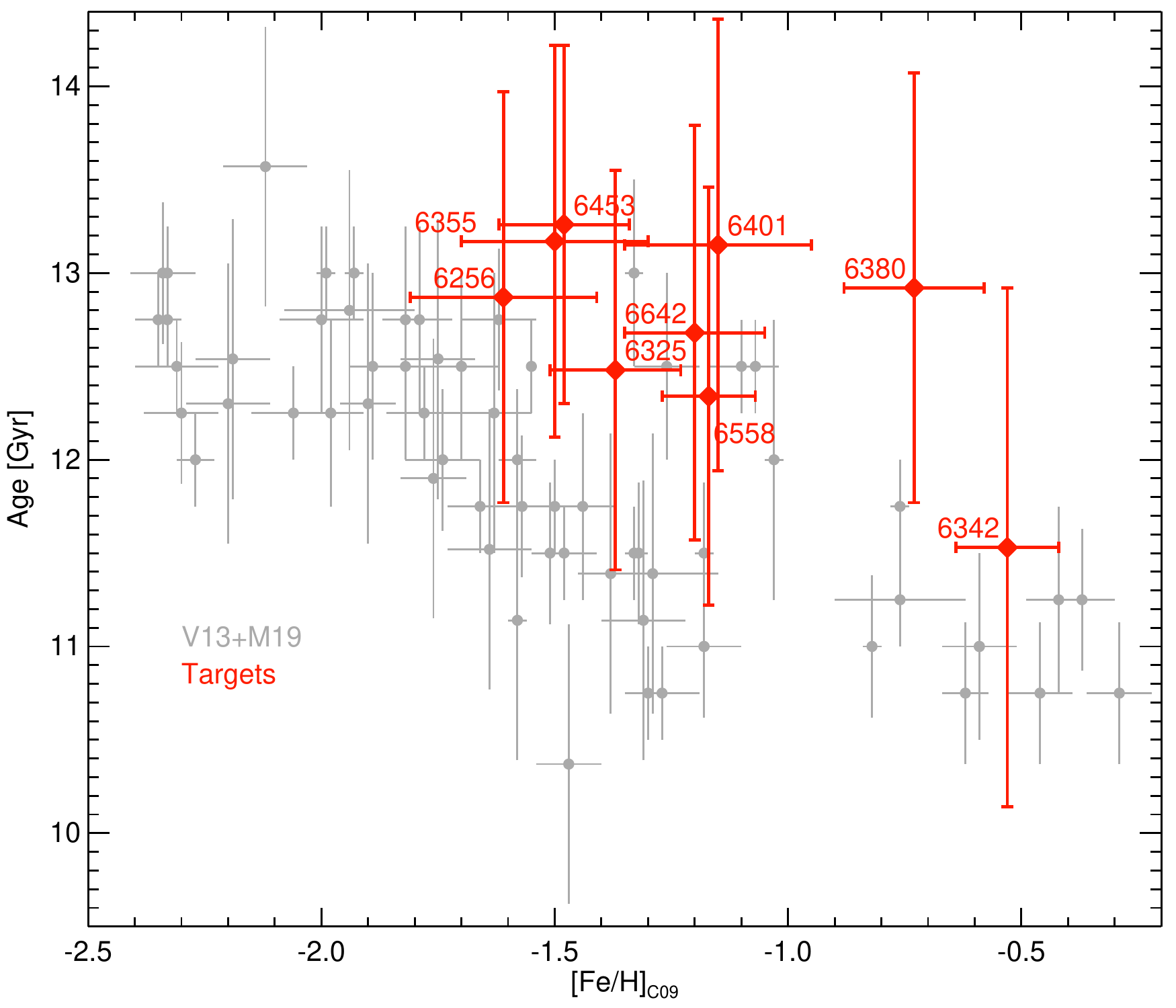}{0.5\textwidth}{}     	 
          \fig{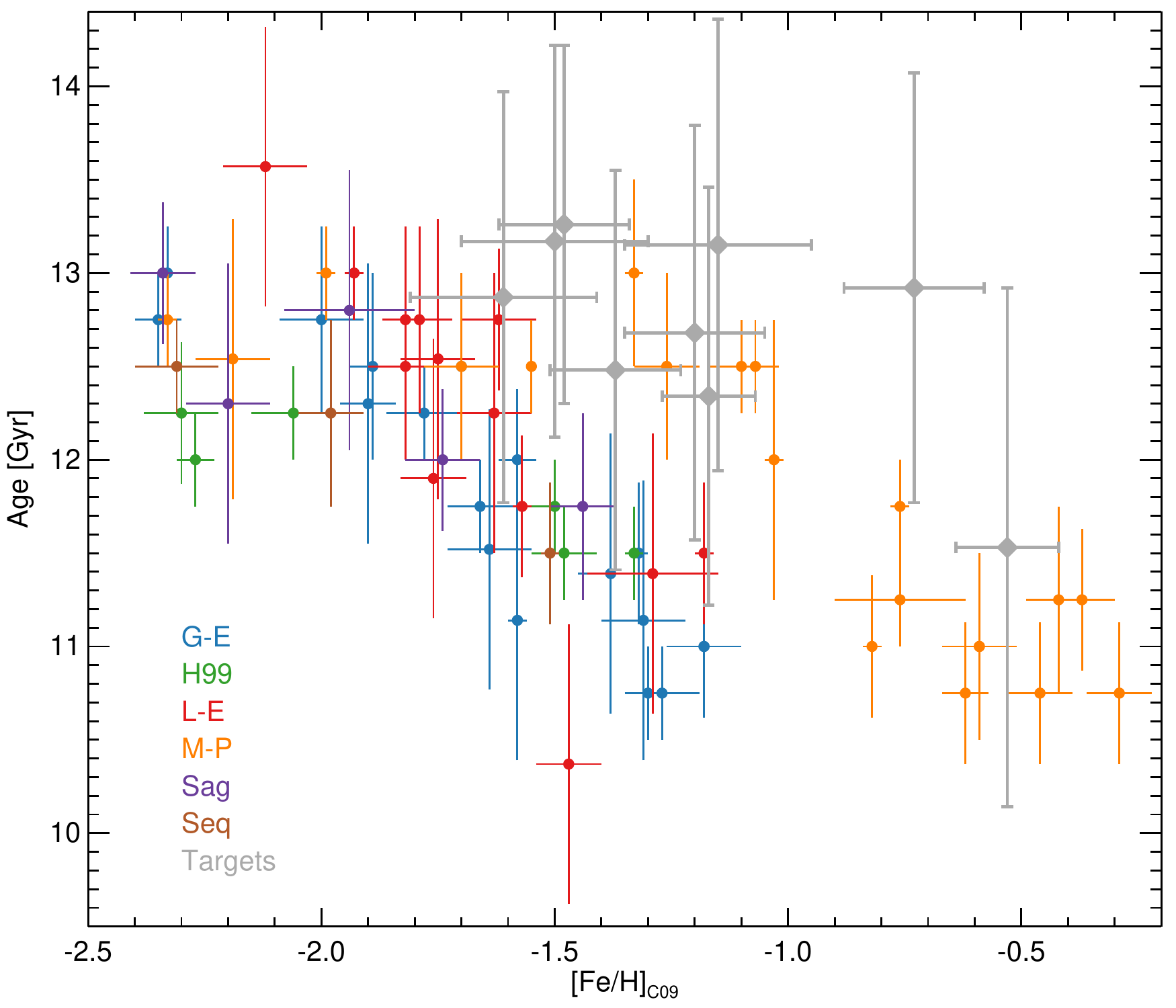}{0.5\textwidth}{}}
\caption{\textbf{Left: } Age-metallicity relation from \citeauthor{v13} supplemented with metal-poor clusters from \citet{fb10} following \citet{massari19} (shown in grey), where horizontal error bars are [Fe/H] uncertainties from \citeauthor{c09}.  Our target clusters are overplotted in red and labeled with their NGC number.  \textbf{Right:} Same, but now literature clusters are color-coded according to their Galactic component from \citet{massari19}, and our target clusters are overplotted using large grey diamonds.
\label{amrfig}}
\end{figure}

To view our target cluster ages in the context of the MWGC age-metallicity relation, in the left panel of Fig.~\ref{amrfig} we show our targets in red, and in grey we show the 55 clusters from \citeauthor{v13}, supplemented by 14 clusters from \citet{fb10} with [Fe/H]$\leq$$-$1.1 following \citet{massari19}.  
The reference cluster ages from \citeauthor{v13} (see Table \ref{refclustab}) have been used to place our target cluster relative ages (Table \ref{htab}) on their MWGC age scale, adding a contribution of 0.3 Gyr (see Sect.~\ref{methtestsect}) to the relative age uncertainties in Table \ref{htab}.
This plot is reproduced in the right-hand panel of Fig.~\ref{amrfig}, but now color-coding the literature clusters by their kinematically-identified progenitor from \citet{massari19}.  
We see that the metal-rich cluster NGC 6342 is fairly young, consistent with other metal-rich MWGCs formed in situ, bolstering its classification from \citet{massari19}.  The remainder of our sample is relatively old, to a remarkably consistent extent: Their mean age is 12.9$\pm$0.4 Gyr, at least as old as any other component of the MWGC age-metallicity relation.  This is in excellent agreement with a mean age of 12.86$\pm$0.36 Gyr for metal-intermediate ($-$1.5$\lesssim$[Fe/H]$\lesssim$$-$0.85) bulge globulars quoted by \citet{oliveira20}, who compiled literature ages to update a similar analysis by \citet{saracino6569}.  However, these age measurements are quite heterogenous, derived using observations in different bandpasses, different age measurement techniques, and in some cases different evolutionary models and different bolometric corrections.  We have attempted to mimimize systematics by keeping all of these ingredients consistent with \citeauthor{v13} to the greatest extent possible to permit a comparison that is reliable at least in a relative sense.  Accordingly, we suggest that applications of heterogenous techniques and observations to measure ages of highly extincted MWGCs ideally should be validated on well-studied low-extinction MWGCs \citep[e.g.][]{correnti_ir1,correnti_ir2} before blindly applying them to MWGCs along heavily extincted sightlines.

Fortuitously, much of our sample lies at [Fe/H] values of $-$1.5$\lesssim$[Fe/H]$\lesssim$$-$1.1, where the discriminatory power is maximized to separate clusters formed in situ, which are older in this metallicity range, from the younger branch of clusters that were accreted from various progenitors.  The old ages of our target clusters in this metallicity range are generally consistent with the kinematic classification of \citet{massari19}, who list seven of our nine target clusters as being formed in situ, with NGC 6401 and NGC 6453 belonging to the low-energy structure dubbed Kraken by \citet{kruijssen_gc2}.  Meanwhile, the location of these two clusters in Fig.~\ref{amrfig} points towards an in situ origin as being perhaps more likely, but especially in light of uncertainties on the chemical abundances of these clusters, membership to Kraken cannot be strongly excluded.  As demonstrated by \citet{massari19} and \citet{kruijssen_gc2}, orbital elements are key to assigning clusters to a progenitor on a case-by-case basis, and these orbital elements rely in turn (in addition to accurate and precise absolute proper motions) on heliocentric distances.  These distances may be particularly uncertain for inner Milky Way globulars due to the variable extinction law towards these sightlines, and just as an example, an assumption of $R_{V}$=2.5 instead of $R_{V}$=3.1 for clusters with $E(B-V)$$~\sim$ 1 at the approximate distance of the Galactic bulge can change their distances by $>$10\% ($>$1 kpc), potentially placing them on the opposite side of the bulge/bar.

Comparing the ages we measure with those in the literature, there are only two clusters with extant age measurements: For NGC 6256, an isochrone-fitting analysis of photometry from the first-epoch \textit{HST} imaging was recently presented by \citet{cadelano6256}, who found an age of 13.0$\pm$0.6 using presumably identical Victoria-Regina isochrones but a different (maximum likelihood) fitting technique,
in excellent agreement with our age of 12.9$\pm$1.0 Gyr (Table \ref{htab}).  
Meanwhile, NGC 6342 is included in the relative age study by \citet{deangeli05}, who find a young relative age, consistent with our results.

Lastly, the location of NGC 6380 in the age-metallicity relation is, thus far, unique.  One possibility is that our assumed metallicity of [Fe/H]$\sim$$-$0.7 is erroneous, since literature estimates of the cluster metallicity vary:
CaII triplet calibrations based on near-IR photometry \citep{m14} as well as near-IR photometric indices (\citeauthor{valenti10}) give values of [Fe/H]$\sim$$-$0.7 to $-$0.8, while earlier spectroscopic indices indicated a higher value of [Fe/H]$\sim$$-$0.4 (\citeauthor{c09}).  However, during the peer review process, elemental abundances based on high-resolution near-IR spectroscopy of 12 RGB members were published by \citet{jose6380}, who report [Fe/H]=$-$0.73$\pm$0.06, consistent with other recent measurements, and they also confirm that this cluster is enhanced in $\alpha$ elements. Nevertheless, as a check on this result, we use three tests:
\begin{enumerate}
  \item We have remeasured the near-IR slope of the upper RGB \citep[][see their sect.~4.4]{cohenvvv} from the same 2MASS-calibrated VVV PSF photometry used in that study, but after applying our differential reddening corrections and selecting members from \citet{gaiamems}, finding a slope of $-$0.102$\pm$0.003, implying [Fe/H]=$-$0.54$\pm$0.09, although the slope-[Fe/H] calibration has a scatter of 0.2 dex. 
\item We measured the difference between the F814W magnitude of the red horizontal branch (RHB) and the magnitude of the red giant branch bump (RGBB), using a maximum likelihood fit of an exponential plus single or double Gaussian to the RGB \citep[e.g.][]{natafbump,cohenamiga}.  For NGC 6380, we find $\Delta I^{\rm RHB}_{\rm RGBB} \approx \Delta F814W^{\rm RHB}_{\rm RGBB}$ = 0.291$\pm$0.021, 
  resulting in [M/H]=$-$0.44$\pm$0.02 for NGC 6380 (corresponding to [Fe/H]=$-$0.73 assuming [$\alpha$/Fe]=+0.4) using eq.~3 of \citet{natafbump}, although they estimate their calibration between $\Delta I^{\rm RHB}_{\rm RGBB}$ and [Fe/H] to have an intrinsic scatter of $\sim$0.05 dex, which they suggest could be due to MWGC age variations.
\item We reperformed our relative age dating procedure for NGC 6380 assuming the higher value of [Fe/H]=$-$0.4 given by \citeauthor{c09}, finding an age well over the age of the Universe, although when assuming [Fe/H]=$-$1.0 for NGC 6380 we find an age of 13.0 Gyr, essentially identical to the result plotted in Fig.~\ref{amrfig}.  This lower value of [Fe/H] (or a somewhat younger age, by $\sim$0.5 Gyr) would bring this cluster well within $\sim$1$\sigma$ consistency with the in situ branch of the MWGC age-metallicity relation, but is at odds with the photometric metallicity indicators above.
  \end{enumerate}
One intriguing possibility to explain the location of NGC 6380 in the age-metallicity plane could be elemental abundance variations that could mimic an older age (see Sect.~\ref{otherchemsect}), but this has been largely ruled out for several potential culprits (O, Mg, Si) by \citet{jose6380}.  However, \citet{jose6380} also point out the unusually large extent of N enhancement in this cluster as well as potential correlations of Ce with N and Al.  High resolution optical spectroscopy and/or blue-ultraviolet photometry could be particularly beneficial in this case to characterize the multiple stellar populations in this cluster and constrain any potential helium enhancement \citep[e.g.][]{milone_dely}.

\subsection{The Impact of $R_{V}$ \label{rvsect}}

We explore the impact of varying the total-to-selective extinction ratio $R_{V}$ by calculating an alternate set of bolometric corrections assuming $R_{V}$=2.5 rather than the classical value of $R_{V}$=3.1.  Extinction towards the Galactic bulge is known to be non-standard (see \citealt{natafrev} for a review), and such low values of $R_{V}$ have been measured towards the bulge \citep{sumi,nataf_rv25}, although the parametrization of the extinction law is still a topic of some debate \citep{nataf_extinction,schlafly16}.  Hence, our intention is not to advocate for a particular $R_{V}$ value, but rather to test if and how a substantial ($\sim$20\%) $R_{V}$ variation impacts derived cluster relative ages.

We assess the impact of a change in $R_{V}$ by examining two hypothetical MWGCs, each with true values of $E(B-V)$=1.5 and $R_{V}$=2.5, and with [Fe/H] values of -0.5 and -1.5 (roughly spanning the range of our target clusters).  In each case, we examine isochrones with ages from 10-14 Gyr in steps of 1 Gyr, and register the isochrones following the usual prescription (Sect.~\ref{techniquesect}).  The registered isochrones corresponding to the correct values of $R_{V}$ and $E(B-V)$ are shown in red in Fig.~\ref{rv_rgbcolfig}, and we now test the impact of the following two erroneous assumptions:

\begin{figure}
\gridline{\fig{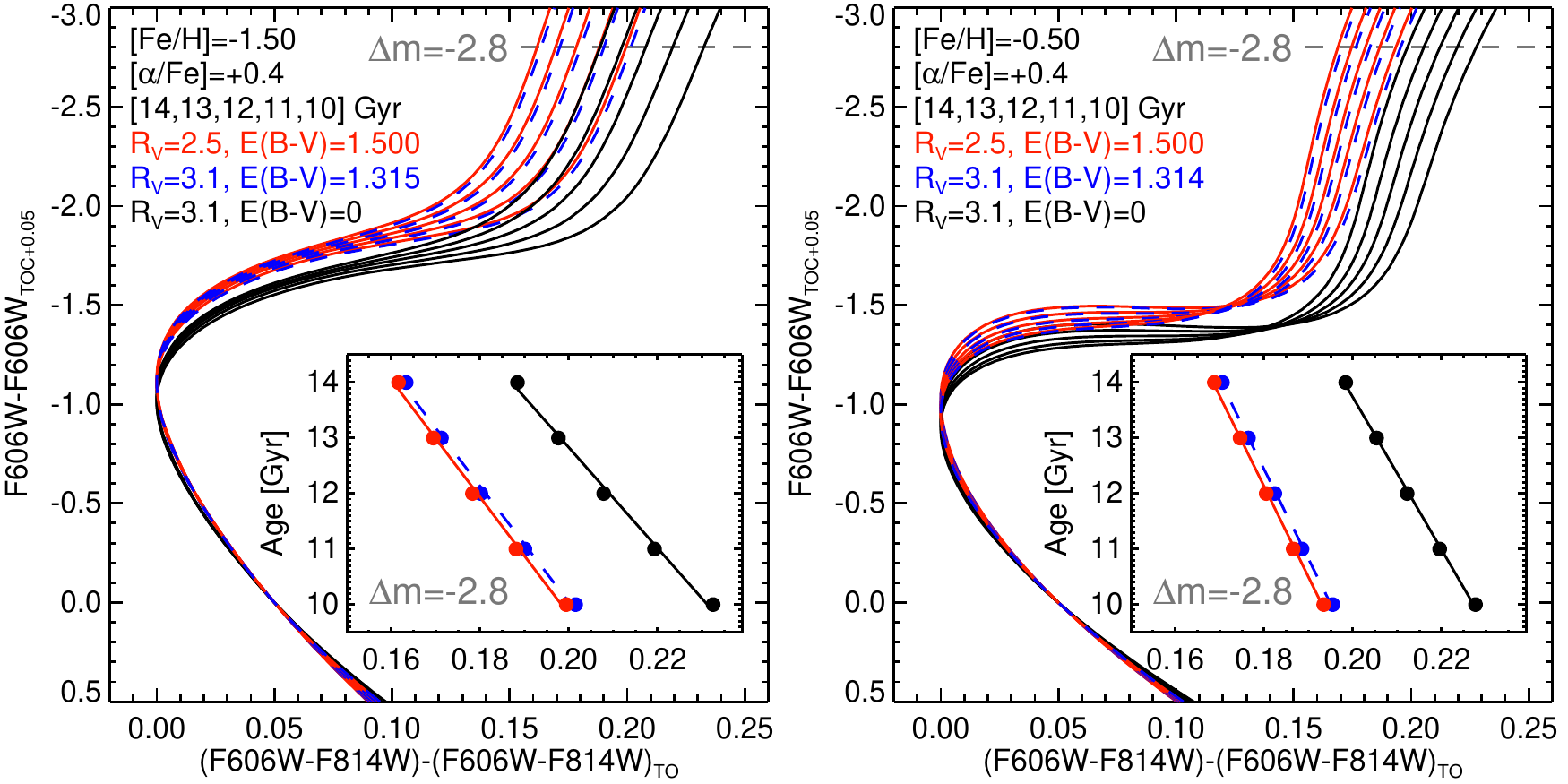}{0.99\textwidth}{}}      	 
\caption{The impact of fixed bolometric corrections or an incorrect value of $R_{V}$ on relative ages, assessed using a hypothetical cluster with true $R_{V}$=2.5, $E(B-V)$=1.5 (at the high end of plausible values for our target clusters; Table \ref{htab}), and [Fe/H]=-1.5 (left panel) or [Fe/H]=-0.5 (right panel), roughly bracketing the range spanned by our target clusters.  In each panel, isochrones from 10-14 Gyr in steps of 1 Gyr are registered following the prescription in Sect.~\ref{relagesect}.  Isochrones that assume the correct value of $R_{V}$=2.5 and $E(B-V)$=1.5 are shown in red, isochrones that incorrectly assume $R_{V}$=3.1 are shown in blue, and isochrones that assume a single, fixed bolometric correction calculated for zero extinction are shown in black.  In the inset in each panel, we show the gradient of RGB relative color, measured at a magnitide of (F606W-F606W$_{TO}$)=$-$2.8 (shown using a grey horizontal line) versus age. \label{rv_rgbcolfig}}
\end{figure}

 First, isochrones that incorrectly assume $R_{V}$=3.1 are shown in blue.  Note that by matching the observed MSTO color of the $R_{V}$=2.5 isochrones, these isochrones correspond to a different value of $E(B-V)$ because of a different assumed $R_{V}$.  Second, isochrones assuming a single fixed zero-extinction value for the bolometric correction (i.e.~not varying as a function of $T_{\rm eff}$, log g, or total extinction) are shown in black.  For each isochrone, following the usual relative age dating procedure, we record the color of the RGB relative to the MSTO at a magnitude of (F606W-F606W$_{\rm TOC+0.05}$)=$-$2.8, shown as a dashed grey horizontal line.  To more clearly illustrate the dependence of this color (as the relative age indicator) on the above assumptions, the inset in each panel shows the gradient of RGB relative color with age, with individual values plotted as filled circles and linear fits as straight lines, revealing the following:
  \begin{itemize}
  \item If a target cluster is at substantially higher $E(B-V)$ than the reference cluster being used, the assumption of fixed bolometric corrections via simply applying horizontal and vertical shifts in their CMDs would result in target cluster ages that are systematically much too old, by $>$2 Gyr or more depending on metallicity for $E(B-V)$$\sim$1.5.  This effect becomes worse at higher metallicities and higher extinction.
  \item When correctly accounting for the variation of bolometric corrections with stellar parameters and total extinction, an incorrect assumption of $R_{V}$ at the level of $\sim$20\% (assuming $R_{V}$=3.1 for a true $R_{V}$=2.5) has only a minor ($<$3\%) impact on the derived ages, resulting in ages which are slightly too young when assuming a value of $R_{V}$ that is too high.  This effect is essentially insignificant for more metal-poor clusters ($<$1\% at [Fe/H]=-1.5 for ages $>$10 Gyr) due to their shallower gradient of relative RGB color versus age.  Accordingly:
   \item Although the \textit{values} of relative ages are drastically misestimated when using a single, fixed bolometric correction in each bandpass, the \textit{gradient} of age as a function of RGB relative color, shown in the insets of Fig.~\ref{rv_rgbcolfig}, is quite insensitive to varying assumptions on $E(B-V)$ or $R_{V}$.  Rather, this gradient depends primarily on metallicity, becoming steeper for more metal-rich clusters (as pointed out by \citeauthor{v13}) and thus worsening the systematic impact of incorrect assumptions on $R_{V}$ and/or $E(B-V)$.
\end{itemize}

The above results all assume a fixed parametrization of the extinction law given by \citet{ccm} and \citet{odonnell}, and further assume that the single parameter $R_{V}$ is capable of capturing its variation.  However, towards the inner Milky Way there is evidence for both a failure of this assumed paramerization as well a multi-parameter dependence of the extinction law \citep{nataf_extinction}.  While beyond the scope of this study, mono-distance and mono-metallicity subsamples such as our proper-motion-selected MWGC members may be ideal to further characterize extinction towards the inner Milky Way and its variation via well-calibrated photometry over a broad range of wavelengths.

\subsection{Helium and Other Chemical Abundance Variations \label{otherchemsect}}

The relative age dating technique we use here is sensitive to variations in elements other than iron, including He, $\alpha$- and light elements such as O, Mg and Si (see e.g.~fig.~17 of \citeauthor{v13}), between any target and reference cluster.  Regarding helium, there are two reasons why a substantial global enhancement is unlikely.  First, we already find relatively old ages ($\sim$13 Gyr to within uncertainties) for all but the young, metal-rich cluster NGC 6342 (which is spectroscopically verified to be a chemical twin of its reference cluster) such that any global He enhancement is restricted to fairly small values ($\Delta$Y$\lesssim$0.04 dex) in order to yield ages younger than the age of the universe.  Second, and relatedly, \citeauthor{v13} point out that increasing the global value of He (unlike some other elements) changes the SGB slope, and we found no evidence for a difference in SGB slopes between our target clusters and their respective comparison clusters\footnote{Importantly, this comparison was made \textit{after} correcting the comparison cluster isochrones to the color excess values of our target clusters using our fully self-consistent bolometric corrections, which does affect the shape of the isochrones, including the SGB.}. Moving on from a global \textit{enhancement} in He, a \textit{spread} in He is a common feature of MWGCs.  Our target clusters have estimated masses in the range 4.5$\lesssim$Log M/M$_{\sun}$$\lesssim$5.5 \citep{baumgardt19}, corresponding to He spreads of $\Delta$Y$\lesssim$0.03 dex \citep[][see their fig.~13]{milone_dely}.  Considering the observational uncertainties on our photometry and differential reddening corrections, our data are generally insufficient to disentangle such a modest He spread, which in any case would not shift our age estimates beyond their uncertainties.

Turning to elements other than He, \citeauthor{v13} demonstrate a systematic impact of $>$1 Gyr on relative ages for significant ($\sim$0.4 dex) differences in some light and $\alpha$ elements.  For this reason, we have intentionally chosen comparison clusters with [$\alpha$/Fe] values similar to those seen for bulge globulars, and these values are consistent with values of [$\alpha$/Fe] and [Mg/Fe] found from low-resolution spectroscopy for five of our nine target clusters (\citeauthor{dias}).  However, more generally, there is also no observational evidence that our target clusters would show noteworthy differences in $\alpha$ or light elements with respect to their comparison clusters.  In addition to the minority (three of nine) of our target clusters that have some of these abundance ratios available from high-resolution optical spectroscopy (Table \ref{targfehtab}), a more general comparison of MWGC elemental abundance ratios over a broad metallicity baseline \citep{carretta_oonly,carretta_elements} does not show a systematic offset from those seen for bulge globulars \citep{j16,puls6366} with respect to these elements.  The only potential exception is Si, for which two metal-intermediate inner Milky Way clusters (NGC 6522 and NGC 6558) show atypically low values close to [Si/Fe]$\sim$0 \citep{barbuy6558} that could mimic an older age (by $\lesssim$1 Gyr).   However, without high-quality spectroscopic data for the majority of our target clusters, we cannot comment further on a case-by-case basis, and self-consistent multi-element abundance ratios are indeed critical to place stringent constraints on MWGC ages.

\section{Summary and Conclusions}

We have derived relative ages for nine inner Milky Way globular clusters, minimizing systematics by using the same filters, evolutionary models, bolometric corrections, and relative age measurement technique as \citeauthor{v13}.  Two independent tests of our methodology indicate that the ages we measure using a classical relative age-dating technique should be directly comparable to those of \citeauthor{v13} at the level of 0.3 Gyr or better.  We find that NGC 6342 is a typical young ($\sim$11.5 Gyr) metal-rich cluster formed in situ, while the remainder of our sample is uniformly very old, with a mean age of 12.9$\pm$0.4 Gyr assuming the \citeauthor{v13} MWGC age scale.  The location of these eight MWGCs in the MWGC age-metallicity plane suggests that most or all of them formed in situ, although more secure kinematics-based classifications require distances to be reliably determined from multi-wavelength measurements of interstellar extinction towards the target clusters and/or direct measurement of parallaxes from future astrometric missions.  We have explored the impact of both assuming a fixed zero-extinction bolometric correction, which drastically shifts ages to higher values, as well as assuming an incorrect value of $R_{V}$, which has a small ($<$3\%) systematic effect on MWGC ages over the range of metallicity (-1.5$\lesssim$[Fe/H]$\lesssim$-0.5) and color excess ($E(B-V)$$<$1.5) we explore.  Variations in elemental abundances other than [Fe/H] could be responsible for the unusual location of NGC 6380 in the MWGC age-metallicity plane, and such measurements from high-resolution spectroscopy on an internally consistent scale could pose more stringent constraints on MWGC relative ages.

\acknowledgements

Support for program HST GO-15065 was provided by NASA through a grant from the
Space Telescope Science Institute, which is operated by the Association of
Universities for Research in Astronomy, Inc., under NASA contract NAS 5-26555. 

LC acknowledges support from the Australian Research Council Future Fellowship FT160100402.

\facilities{HST (ACS/WFC,WFC3/UVIS)}




\end{document}